\documentclass[12pt,fleqn]{article}
\usepackage{amsmath,amssymb}
\usepackage{bm}
\usepackage[dvips]{graphicx}
\newcommand{\re}{\text{Re}}

\newcommand{\Tr}{\text{Tr}}

\allowdisplaybreaks[3]

\begin{document}
\title{Photoproduction of $K^*$ for the study of $\Lambda(1405)$}
\author{T.~Hyodo$^a$\footnote{E-mail :
hyodo@rcnp.osaka-u.ac.jp}, A.~Hosaka$^a$, 
M.~J.~Vicente~Vacas$^b$ and E.~Oset$^b$}
\date{\today}
\maketitle 

\begin{center}
$^a$\textit{Research Center for Nuclear Physics (RCNP),
Ibaraki, Osaka 567-0047, Japan} \\
$^b$\textit{Departmento de F\'isica Te\'orica and IFIC,
Centro Mixto Universidad de Valencia-CSIC,
Institutos de Investigaci\'on de Paterna, Aptd. 22085, 46071
Valencia, Spain}
\end{center}

\vspace{0.4cm}

\abstract{ The photo-induced $K^*$ vector meson production is  investigated for
the study of the $\Lambda(1405)$ resonance.   This reaction is particularly
suited to the isolation of the second pole in the $\Lambda(1405)$ region which
couples  dominantly to the $\bar K N$ channel.  We obtain the mass distribution
of the $\Lambda(1405)$  which peaks at 1420 MeV, and differs from the nominal
one. Combined with several other reactions, like the $\pi^- p \to K^0 \pi
\Sigma$ which favours  the first pole, this detailed study   will reveal a
novel structure of the $\Lambda(1405)$ state.   }


\vspace*{1cm}


In recent years, we have been observing a remarkable development 
in hadron physics, especially in baryon resonances.  
Chiral models which implement strong $s$-wave meson-baryon interactions
have been showing that some of the $1/2^-$ resonances,
such as the $\Lambda(1405)$, are strongly dominated by quasi-bound
states of coupled meson-baryon 
channels~\cite{Kaiser:1995eg,Kaiser:1997js,Oset:1998it,Lutz:2001yb}.
The case of the $\Lambda(1405)$ as a quasibound state is not a merit of the
chiral Lagrangians since it was previously obtained in a unitary coupled channel
approach in Refs.~\cite{annphys10.307,PR153.1617,Jones:1977yk}.
The use of chiral Lagrangians
has allowed a systematic approach to face the meson-baryon interaction.
In these pictures, the resonances generated may be regarded as another 
realization of  
5-quark dominated states, although the lowest Fock space  
to generate
the quantum numbers of $\Lambda$
starts from 3-quark states. The $s$-wave meson-baryon
interaction at lowest order
in the chiral Lagrangians is given by 
the Weinberg-Tomozawa term which contains an attractive interaction 
in the $\bar K N$ channel as well as its couplings to other 
meson-baryon channels. Therefore,
confirmation of this picture is important in order to understand 
better the non-perturbative dynamics of QCD.

One interesting finding concerns the structure of the 
$\Lambda(1405)$ resonance; 
several groups have reported that there are two poles 
in the region of $\Lambda(1405)$ in analyses based on 
the chiral unitary models~\cite{Oller:2000fj,Oset:2001cn,Jido:2002yz,
Garcia-Recio:2002td,
Hyodo:2002pk,Hyodo:2003qa,Garcia-Recio:2003ks,Nam:2003ch}.  
The existence of two poles was first found  in the context of the 
cloudy bag model~\cite{Fink:1990uk}, and recent studies
of chiral dynamics reveal the detailed structure of these poles.
For instance, in Ref.~\cite{Jido:2003cb}, they found two poles at
$z_1 = 1390 - 66i$ MeV and 
$z_2 = 1426 - 16i$; 
the former at lower energy and with a wider width 
couples dominantly to $\pi \Sigma$ channels, 
while the latter at higher energy with a narrower width 
couples dominantly to $\bar K N$ channels.  
If this is the case,
the form of the invariant mass distribution of $\pi\Sigma$, where
the $\Lambda(1405)$ is seen,
depends on the particular reaction used to generate 
the $\Lambda(1405)$~\cite{Jido:2003cb}.
In fact, different shapes of mass distributions
were observed in previous theoretical 
studies~\cite{Nacher:1998mi,Nacher:1999ni}
and in Ref.~\cite{Hyodo:2003jw} it was found that the $\pi^- p \to K^0 \pi
\Sigma$ reaction was particularly selective of the first $\Lambda$ pole. 
It is therefore desirable to study the nature of 
the $\Lambda(1405)$ focusing on whether such two poles 
really exist in the nominated resonance region.

In this paper, we propose another, hopefully better, 
reaction induced by 
photons for the extraction of the second pole around the 
$\Lambda(1405)$ resonance: 
$\gamma p \to K^* \Lambda(1405)\to \pi K \pi \Sigma$.  
A great advantage of this reaction is the use of 
a linearly polarized photon beam and the observation of the
angular distribution of $\pi K$ decaying from $K^*$, 
which is correlated with the linear polarization of the
photon.

\begin{figure}[tbp]
    \centering
    \includegraphics[width=8cm,clip]{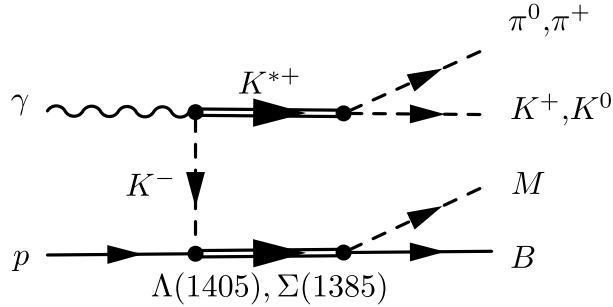}
    \caption{\label{Lambdareaction}
    Feynman diagram for the reaction.  
    $M$ and $B$ denote
    the meson and baryon of
    ten coupled channels of $S=-1$ meson-baryon scattering.
    In this paper we only take $\pi\Sigma$ and $\pi\Lambda$ channels
    into account.}
\end{figure}

As we shall see below, if we produce a $\pi K$ system in a plane perpendicular
to the photon polarization, 
 the $t$-channel exchanged particle 
is dominated by the kaon; heavier strange mesons contributions 
should be suppressed due to their larger masses.  
Ignoring (hopefully small) background contributions from, 
for instance, 
unknown higher nucleon resonances also, 
it is sufficient to consider only the processes 
with kaon exchange,
as shown in Fig.~\ref{Lambdareaction}.  
The exchanged kaon rescatters in isospin $I = 0$ and 1 channels.  
Then the former couples strongly to  $\Lambda(1405)$, 
especially to the higher pole, while the latter does it to $\Sigma(1385)$.  
We utilize the $s$-wave meson-baryon scattering amplitude
calculated by the chiral unitary model~\cite{Oset:2001cn,Hyodo:2002pk}.
The amplitude generates the $\Lambda(1405)$ resonance
dynamically, 
while the $\Sigma(1385)$ is not generated because it is
a $p$-wave resonance. In order to perform a realistic calculation,
we introduce the $\Sigma(1385)$ field explicitly.

\begin{figure}[tbp]
    \centering
    \includegraphics[width=7cm,clip]{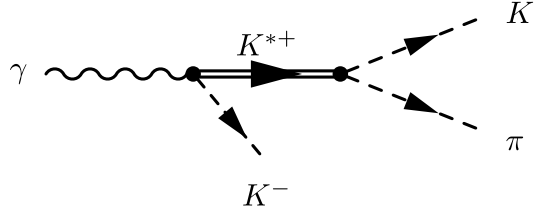}
    \caption{\label{fig:vector}
    Feynman diagram for $\gamma\to K^-K\pi$.}
\end{figure}

The scattering amplitude as described by the 
diagram of Fig.~\ref{Lambdareaction} can be divided into two
parts
\begin{equation}
    -it=(-it_{\gamma\to K^-K\pi})\frac{i}{p_{K^-}^2-m_{K^-}^2}
    (-it_{K^-p\to MB}) \ .
    \label{eq:totamp}
\end{equation}
The former part $(-it_{\gamma\to K^-K\pi})$,
as shown in Fig.~\ref{fig:vector},
is derived from the following 
effective Lagrangians~\cite{Oh:2003kw,Palomar:2002hk}
\begin{align}
    \mathcal{L}_{K^*K\gamma}
    &= g_{K^*K\gamma}\epsilon^{\mu\nu\alpha\beta}
    \partial_{\mu}A_{\nu}(\partial_{\alpha}K^{*-}_{\beta}K^+
    +\partial_{\alpha}\bar{K}^{*0}_{\beta}K^0)
    +\text{h.c.} \ ,
    \label{KstarKg} \\
    \mathcal{L}_{VPP}
    &=-\frac{ig_{VPP}}{\sqrt{2}}
    \Tr \bigl(
    V^{\mu}[\partial_{\mu}P,P]\bigr) \ ,
    \label{KstarKpi}
\end{align}
where $K$, $A_{\mu}$, $V_\mu$ and $P$ are the kaon, 
photon, octet vector meson and octet 
pseudoscalar meson fields, respectively.
The coupling constants are determined from the empirical
partial decay width of $K^*$:  
$\Gamma_{K^{*\pm}\to K^{\pm}\gamma}=0.05$ MeV
and $\Gamma_{K^{*\pm}\to K\pi}= 51$ MeV. 
The resulting values are 
$
|g_{\gamma K^{*\pm}K^{\pm}}|= 0.252 \;\;[\text{GeV}^{-1}]\
$
and 
$g_{VPP}=-6.05$.  
The latter $g_{VPP}$ is the universal vector meson coupling 
constant. 
We note that the effective Lagrangian \eqref{KstarKg} is  
consistent with a vector meson dominance model~\cite{Bramon:1992kr}.
Using the above interaction Lagrangians, 
the amplitude for 
$\gamma \to K^{*+} K^- \to K^0 \pi^+ K^-$ is given 
by~\footnote{
For the final state $K^+ \pi^0$, the amplitude is reduced by factor
$1/\sqrt{2}$, and therefore, the resulting cross section becomes one half.
In the rest of this paper, we show the result for $K^0 \pi^+$.}
\begin{equation}
    -it_{\gamma\to K^-K^0\pi^+}
    = \frac{i
    \sqrt{2}g_{VPP}
    \epsilon^{\mu\nu\alpha\beta}p_{\mu}(K^0)p_{\nu}(\pi^+)
    k_{\alpha}(\gamma)\epsilon_{\beta}(\gamma)}
    {P_{K^*}^2-M_{K^*}^2+iM_{K^*}\Gamma_{K^*}}
    \ , 
    \label{eq:vecamp}
\end{equation}
where $p$ and $k$ are the momenta of the particle in parentheses,
$\epsilon_{\mu}(\gamma)$ is the polarization vector of photon, and 
$\Gamma_{K^*}$ is the total decay width of $K^*$, for which 
we use the energy dependent one for a virtual $K^*$, 
$\Gamma_{K^*} =A p_{CM}^3$, 
where 
$p_{CM}$ is the two-body relative momenta of the final state, and
$A=2.05\times10^{-6}[\text{MeV}^{-2}]$ 
such that $\Gamma_{K^*}\sim 51$ MeV at the
resonance position. 
Eq. \eqref{eq:vecamp} is instructive to show the correlations between the photon
polarization and the $K^0$ and $\pi^+$ momenta. In order to maximize the
contribution of the t-channel we select the $K^*$ in the direction of the
photon. Then, it is easy to see that the amplitude is proportional to $\sin\phi$
where $\phi$ is the angle between the plane defined by the  $K^0$ and $\pi^+$ 
momenta and the photon polarization (in the Coulomb gauge, $\epsilon^0=0$).
Hence, the maximum strength of the amplitude occurs when this plane is
perpendicular to the photon polarization.

In addition one needs not to worry about symmetrization in the case where there
are two equal charge pions in the final state. In this case the interference
term is zero and one can omit the symmetrization and the $1/2$ factor in the
cross section.
%

\begin{figure}[tbp]
    \centering
    \includegraphics[width=12cm,clip]{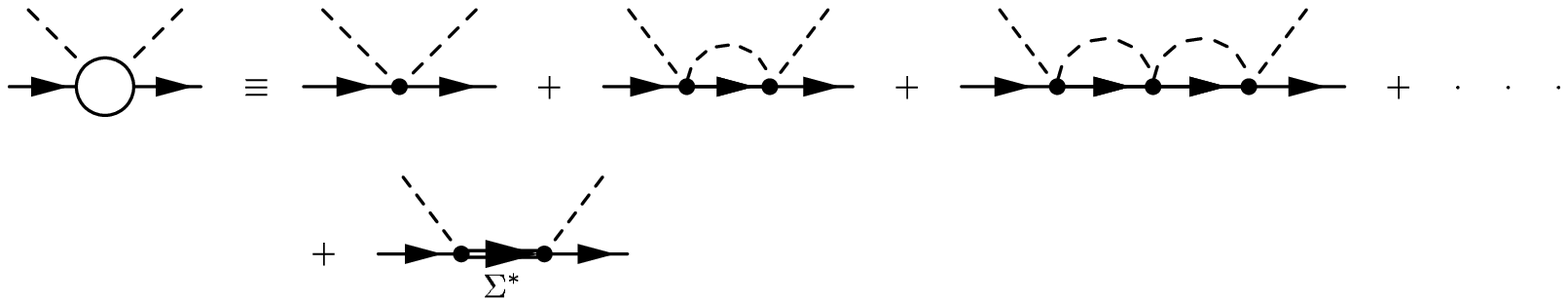}
    \caption{\label{fig:Baryon}
    Feynman diagram for $ K^- p \to MB$.}
\end{figure}

The amplitude $(-it_{K^-p\to MB})$ consists of two parts,
as shown in Fig.~\ref{fig:Baryon}
\begin{equation}
    -it_{K^-p\to MB}(M_I)
    =-it_{ChU}(M_I)-it_{\Sigma^*}(M_I) \ ,
    \label{eq:baryonamp}
\end{equation}
where $-it_{ChU}$ is the meson-baryon scattering amplitude
derived from the chiral unitary model, and $-it_{\Sigma^*}$
is the $\Sigma(1385)$ pole term.
$M_I$ is the invariant mass for 
$K^-p$ system, which is determined by 
$M_I^2=(p_\gamma+p_N-p_{K^*})^2 $.
In the chiral unitary model~\cite{Oset:2001cn,Hyodo:2002pk}, 
the coupled channel amplitudes are obtained by
\begin{equation}
    t_{ChU}(M_I)=[1-VG]^{-1}V\, , 
    \label{eq:ChUamp}
\end{equation}
where $G$ is the meson-baryon loop function and $V$ is the
kernel interaction derived from the Weinberg-Tomozawa term
of the chiral Lagrangian. This amplitude  reproduces well the total cross
sections for several channels. It also leads to dynamically generated 
resonances in good agreement with experiment.
Since the $\Sigma(1385)$ is not generated in this resummation because
it is a $p$-wave resonance, we introduce it explicitly with its coupling 
to channel $i$ ($\Sigma(1385) \to MB$) 
which is deduced from the $\pi N \Delta$ using $SU(6)$ symmetry 
in~\cite{Oset:2000eg,Jido:2002zk} and given  by
\begin{equation}
    -it_{\Sigma^* i}
    =c_i\frac{12}{5}\frac{g_A}{2f}\bm{S}\cdot \bm{k}_i \ ,
    \label{eq:Sigmacouple}
\end{equation}
where 
$g_A=1.26$,
and we use the meson decay constant $f=93\times 1.123$ 
MeV~\cite{Oset:2001cn}.
This is a nonrelativistic 
form for the transition between spin 1/2 and 3/2 particles, 
where $\bm{S}$ is a spin transition operator~\cite{Brown:1975di}
and the coefficients $c_i$ are given in Table~\ref{tbl:1}.
Note that these couplings reproduce well the observed branching 
ratio of $\Sigma(1385)$ decay into $\pi \Lambda$ and 
$\pi \Sigma$.  Then we have the amplitude
\begin{equation}
    -it_{\Sigma^*}(M_I)
    =- c_1c_i
    \left(\frac{12}{5}\frac{g_A}{2f}\right)^2
    \bm{S}\cdot \bm{k}_1
    \bm{S}^{\dag}\cdot \bm{k}_i
    \frac{i}{M_I-M_{\Sigma^*}+i\Gamma_{\Sigma^*}/2}
    F_f(k_1) \ ,
    \label{eq:Sigmaamp}
\end{equation}
where we have introduced a strong form factor 
$F_f(k_1)$ for the vertex $K^-p\Sigma^*$
in order to account for the finite size structure of the baryons.
We adopt a monopole type 
$F_f(q) = (\Lambda^2 - m_K^2)/(\Lambda^2 - q^2)$ with 
$\Lambda=1$ GeV.
In the present reaction around the region of $\Lambda(1405)$,
the effect of the form factor is not very large.

\begin{table}[tbp]
    \centering
    \caption{$c_i$ coefficients}
    \begin{tabular}{|c|c|c|c|c|c|c|c|c|c|c|}
	\hline
	channel $i$ & $K^-p$ & $\bar{K}^0n$ & $\pi^0\Lambda$
	& $\pi^0\Sigma^0$ & $\eta\Lambda$ & $\eta\Sigma^0$
	& $\pi^+\Sigma^-$ & $\pi^-\Sigma^+$ & $K^+\Xi^-$ & $K^0\Xi^0$  \\
	\hline
	$c_i$ & $-\sqrt{\frac{1}{12}}$ & $\sqrt{\frac{1}{12}}$
	& $\sqrt{\frac{1}{4}}$ & 0 & 0 & $-\sqrt{\frac{1}{4}}$
	& $-\sqrt{\frac{1}{12}}$ & $\sqrt{\frac{1}{12}}$
	& $\sqrt{\frac{1}{12}}$ & $-\sqrt{\frac{1}{12}}$  \\
	\hline
    \end{tabular}
    \label{tbl:1}
\end{table}

The cross section is then given by the squared 
amplitude of Eq.~\eqref{eq:totamp}
integrated over the four-body phase space.  
After eliminating four momenta variables from the twelve 
momenta (of four particles), we can write a total 
cross section as a function of the incident energy $\sqrt{s}$:
\begin{equation}
    \sigma(\sqrt{s})
    =\frac{2MM_{\Sigma}}{s-M^2}
    \int \frac{d^3p_1}{(2\pi)^3}\frac{1}{2\omega_1}
    \int \frac{d^3p_2}{(2\pi)^3}\frac{1}{2\omega_2}
    \frac{1}{2}\int_{-1}^{1}d\cos\theta
    \frac{1}{4\pi}\frac{\tilde{P}_3}{M_I} |t(\cos\theta)|^2
    \label{eq:cross}
\end{equation}
where $p_{1(2)}$ and $\omega_{1(2)}$ are the momenta and energy
of the final $K(\pi)$ from $K^*$, and $\tilde{P}_3$ is the relative 
three momentum of $MB$ ($\sim \pi \Sigma$ or $\pi \Lambda$) 
in their center of mass frame.  
The angle $\theta$ denotes the relative angle of $MB$ in the CM frame 
of the total system.   
We have performed this integration by the Monte-Carlo method.
As we have mentioned before, the advantage of this reaction 
is that the identification of the $K^*$ production is cleanly 
done in experiments.  
Observation of the three pions in the process
$K^{*+} \to \pi^+ + K^0 \to \pi^+ + (\pi^+ \pi^-)$
can be made with high efficiency and with all three momenta
measured.

Before going to the numerical results,
here we mention the $MB$ channels decaying from 
the intermediate baryonic state ($B^* \sim \Lambda(1405), \Sigma(1385)$).
There are four possible $MB$ channels 
as shown in Table~\ref{decaychannel}, 
two charged and two neutral channels.  
In the present case, since we have the $K^-p$ channel initially,
the $I=2$ component of $\pi\Sigma$ channel is not allowed.
Considering the Clebsh-Gordan coefficients~\cite{Nacher:1998mi},
the charged channels ($\pi^{\pm}\Sigma^{\mp}$) are from the decay 
of both $\Lambda(1405) (I = 0)$
and $\Sigma(1385) (I = 1)$, while the neutral channels are 
from either one of the two;  
$\pi^0 \Sigma^0$ is from $\Lambda(1405)$ and 
$\pi^0 \Lambda$ is from $\Sigma(1385)$.  

\begin{table}[tbp]
    \centering
    \caption{Possible decay channels from baryons}
    \begin{tabular}{c | c}
    \hline
	Intermediate baryon & Decay channels\\
	\hline
	$\Lambda(1405)$ $I = 0$ & 
	       $\pi^{\pm} \Sigma^{\mp}$, $\pi^0 \Sigma^0$ \\		   
	$\Sigma(1385)$ $I = 1$ & 
	       $\pi^{\pm} \Sigma^{\mp}$, $\pi^0 \Lambda$ \\
		\hline
    \end{tabular}
    \label{decaychannel}
\end{table}

\begin{figure}[tbp]
    \centering
    \includegraphics[width=10cm,clip]{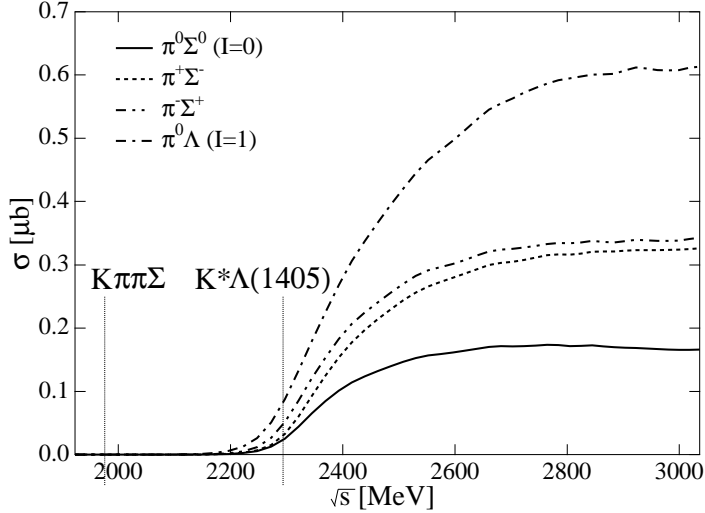}
    \caption{Total cross sections of the process
    with the final states $\pi^0\Sigma^0$ (Solid),
    $\pi^{+}\Sigma^{-}$ (Dashed),
    $\pi^{-}\Sigma^{+}$ (Dash-dot-dotted)
    and $\pi^0\Lambda$ (Dash-dotted)
    in units of [$\mu$b].
    Solid bars indicate the threshold energy of channels.}
    \label{fig:cross}
\end{figure}

Now we present numerical results for total cross sections.  
Unless we observe angular distributions, 
there is not distinction between cross sections of 
polarized and unpolarized processes.  
Therefore, our predictions below are compared with the results 
of both polarized and unpolarized experiments directly.
However, from the experimental point of view it is most practical to concentrate
in the region where the $K^0 \pi^+$ reaction plane is perpendicular to the
photon polarization to maximize the weight of the $K^*$ production mechanism and
reduce possible backgrounds.
In Fig.~\ref{fig:cross}, we show the total cross sections
$\sigma(\gamma + p \to K^* + B^* \to \pi^+K^0 + MB)$ as 
functions of $\sqrt{s}$ for different $MB$ channels.  
As seen in the figure, 
the present mechanism shows up strength at an energy 
slightly lower than the threshold of 
$K^* \Lambda(1405) \sim K^* \Sigma(1385)$ since the 
physical resonances have a finite width and hence a mass distribution.  
In the total cross section, the isospin one ($I=1$)
$MB = \pi^0\Lambda$ channel 
is the largest in size, coming from $B^* = \Sigma(1385)$.  
This might disturb the contribution from $\Lambda(1405)$ of 
$I =0$, unless the separation of these two channels is done.  
However, it  turns out that 
the observation of another charged $\pi$ from the 
intermediate baryon (either $\Lambda(1405)$ or 
$\Sigma(1385)$) helps.  

\begin{figure}[tbp]
    \centering
    \includegraphics[width=10cm,clip]{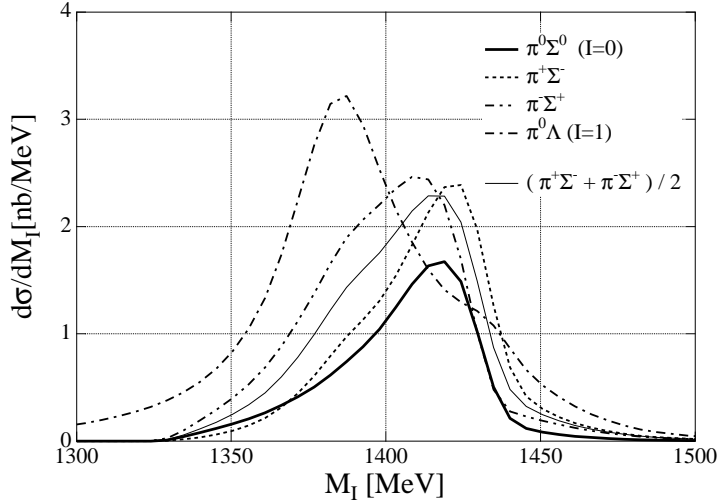}
    \caption{Invariant mass distributions
    of $\pi^0\Sigma^0$ (Thick solid),
    $\pi^{+}\Sigma^{-}$ (Dashed),
    $\pi^{-}\Sigma^{+}$ (Dash-dot-dotted),
    $\pi^0\Lambda$ (Dash-dotted)
    and $(\pi^{+}\Sigma^{-}+\pi^{+}\Sigma^{-})/2$ 
    (Thin solid)
    in units of [nb/MeV].
    Initial photon energy in Lab. frame is 2500 MeV 
    ($\sqrt{s}\sim
    2350$ MeV, 
    threshold of $K^*\Lambda(1405)$).}
    \label{fig:Mdist}
\end{figure}

In order to see this situation, we show in Fig.~\ref{fig:Mdist}
the invariant mass distributions for different decay channels.  
In the figure the initial photon energy is chosen at  
$E_\gamma = 2500$ MeV (the threshold for $K^* \Lambda(1405)$ 
production is 2350 MeV).  
Forgetting about the experimental feasibility, the would-be observable 
in the neutral channel 
is most helpful in order to distinguish the 
contributions from $\Lambda(1405)$ and $\Sigma(1385)$.  
As expected, the $\pi^0 \Sigma^0$ distribution decaying 
from $\Lambda(1405)$ (solid line) has a peak 
around 1420 MeV which is the position of the higher pole.  
In contrast, the $\pi^0 \Lambda$ distribution (dot-dashed line) 
has clearly 
a peak around 1385 MeV, with a 
larger value than the 
$\pi^0 \Sigma^0$ distribution.   
In experiments, the charged states may be observed, which 
contain both $\Lambda(1405)$ and $\Sigma(1385)$ contributions.  
Hence, we show the distribution of charged states by the dashed 
and dash-dot-dotted lines.
The shapes of the three $\pi\Sigma$ distributions have a similar
tendency as the Kaon photoproduction process~\cite{Nacher:1998mi},
which has been confirmed in experiments~\cite{Ahn:2003mv}.
Note also that the contributions from $\Sigma(1385)$ seem to be small
for these channels.

It is worth showing the isospin
decomposition of the distributions of charged states~\cite{Nacher:1998mi}
\begin{equation}
    \frac{d\sigma(\pi^{\pm}\Sigma^{\mp})}{dM_I}
    \propto \frac{1}{3}|T^{(0)}|^2+\frac{1}{2}|T^{(1)}|^2
    \pm\frac{2}{\sqrt{6}}\re (T^{(0)}T^{(1)*}) \ ;\quad
    \frac{d\sigma(\pi^{0}\Sigma^{0})}{dM_I}
    \propto \frac{1}{3}|T^{(0)}|^2 \ ,
    \label{eq:isodecomp}
\end{equation}
where $T^{(I)}$ is the amplitude with isospin $I$.
The factor $1/2$ in front of $|T^{(1)}|^2$
and the ratio of the couplings
$g_{\Sigma^*\pi^{\pm}\Sigma^{\mp}}^2/g_{\Sigma^*\pi^0\Lambda}^2=1/3$
(see Table.~\ref{tbl:1})
explain why the $\Sigma(1385)$ does not affect
the charged $\pi\Sigma$ channels very much,
as compared with the  $\pi\Lambda$ final state.
The difference between $\pi^+\Sigma^-$ and $\pi^-\Sigma^+$ 
comes from the crossed term $\re (T^{(0)}T^{(1)*})$,
and when we sum up the two distributions this term vanishes.
We also show the result for the sum of the charged $\pi\Sigma$ channels
in Fig.~\ref{fig:Mdist} (thin solid line).
The feature that the initial $K^- p$ couples dominantly to the second 
pole of the $\Lambda(1405)$ is well preserved in the total mass 
distribution, although the width of this distribution is slightly larger 
than that of the I=0 resonance because it contains some contribution 
from the $\Sigma(1385)$.  
This is a nice feature and suggests that by observing 
the mass distributions of 
the charged state from the intermediate baryon, 
it would be possible to study the nature of the 
second pole of the $\Lambda(1405)$ resonance.

It is also interesting to see the $I=1$ $s$-wave amplitude
in this energy region,
where the existence of another pole is 
discussed~\cite{Oller:2000fj,Jido:2003cb}.
It was shown in Ref.~\cite{Jido:2003cb} that in the $SU(3)$
decomposition of the meson baryon states the interaction was attractive in a
singlet and two octets, hence 
it is natural to expect the existence of another
$s$-wave $I=1$ resonance in addition to the $\Sigma(1620)$ already 
reported in Ref.~\cite{Oset:2001cn}.
Indeed, a pole is found at $1410 - 40i$
MeV in the model of Ref.~\cite{Oller:2000fj}.
However, the properties of this $I =1$ pole are very sensitive to 
the details of the model since in different models or approximations it appears
in different Riemann sheets, but there is still some reflection on the
amplitudes in all cases.
Therefore, investigation of the $I=1$ $s$-wave amplitude
would bring further information of resonance properties.

\begin{figure}[tbp]
    \centering
    \includegraphics[width=10cm,clip]{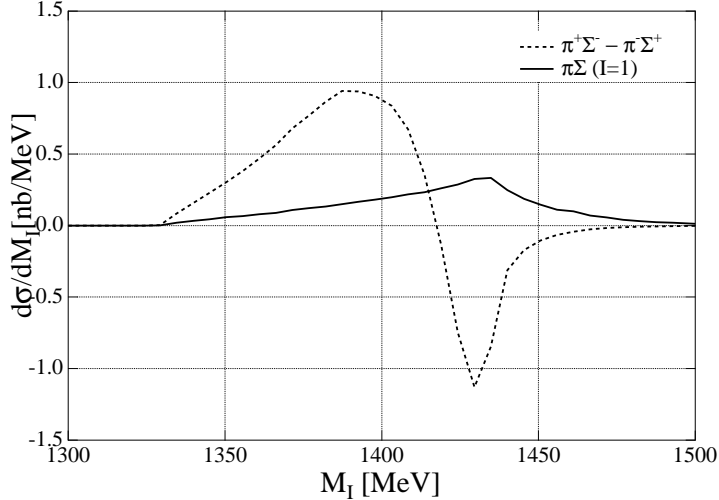}
    \caption{
    Invariant mass distributions
    of $\pi^{+}\Sigma^{-}-\pi^{-}\Sigma^{+}$ (Dashed),
    and $s$-wave, $\pi\Sigma(I=1)$ (Solid),
    in units of [nb/MeV].}
    \label{fig:I0amp}
\end{figure}

We could have the $I=1$ amplitude by combining the three $\pi\Sigma$
channels (see Eq.~\eqref{eq:isodecomp}). However, the $|T^{(1)}|^2$ term 
will contain contributions both from $s$ and $p$-wave, although the
contribution of the $p$-wave to the $\pi\Sigma$ channels is small.
In order to extract the  $I=1$ $s$-wave amplitude we separate 
the T-matrix  into partial waves as
\begin{equation}
    T^{(0)}=T^{(0)}_s \ , \quad
    T^{(1)}=T^{(1)}_s+T^{(1)}_p \ .
    \label{eq:partial}
\end{equation}
Since we are looking at the cross sections where the angle
variable among $MB$ is integrated, the product of
$s$- and $p$-wave
amplitude vanishes.
Then, the difference of the distributions for the  two charged states 
contains only the $T^{(1)}_s$ amplitude
\begin{equation}
    \frac{d\sigma(\pi^+\Sigma^-)}{dM_I}
    -\frac{d\sigma(\pi^-\Sigma^+)}{dM_I}
    =\frac{4}{\sqrt{6}}\re (T^{(0)}_s
    (T^{(1)}_s)^*) \ .
    \label{eq:diff}
\end{equation}
We plot this magnitude in Fig.~\ref{fig:I0amp} with a dashed line.
In principle, it is possible to extract $T^{(1)}_s$ from this quantity
and the distribution of $s$-wave $I=0$ (for instance, from the $\pi^0
\Sigma^0$), parametrizing conveniently the $T^{(0)}_s$ amplitude.
Theoretically, in the present framework, we can calculate the pure
$s$-wave $I=1$ by switching off the $\Sigma(1385)$ and making the
combination of $\pi \Sigma$ amplitudes
\begin{equation}
    \frac{d\sigma(\pi^+\Sigma^-)}{dM_I}
    +\frac{d\sigma(\pi^-\Sigma^+)}{dM_I}    
    -2\frac{d\sigma(\pi^0\Sigma^0)}{dM_I} \ .
     \label{eq:diffmanolo}
\end{equation}
 The results
are shown in Fig.~\ref{fig:I0amp} (Solid line) and a small peak is seen as
a reflection of the approximate resonant structure  predicted in 
Refs.~\cite{Oller:2000fj,Jido:2003cb}.

\begin{figure}[tbp]
    \centering
    \includegraphics[width=8cm,clip]{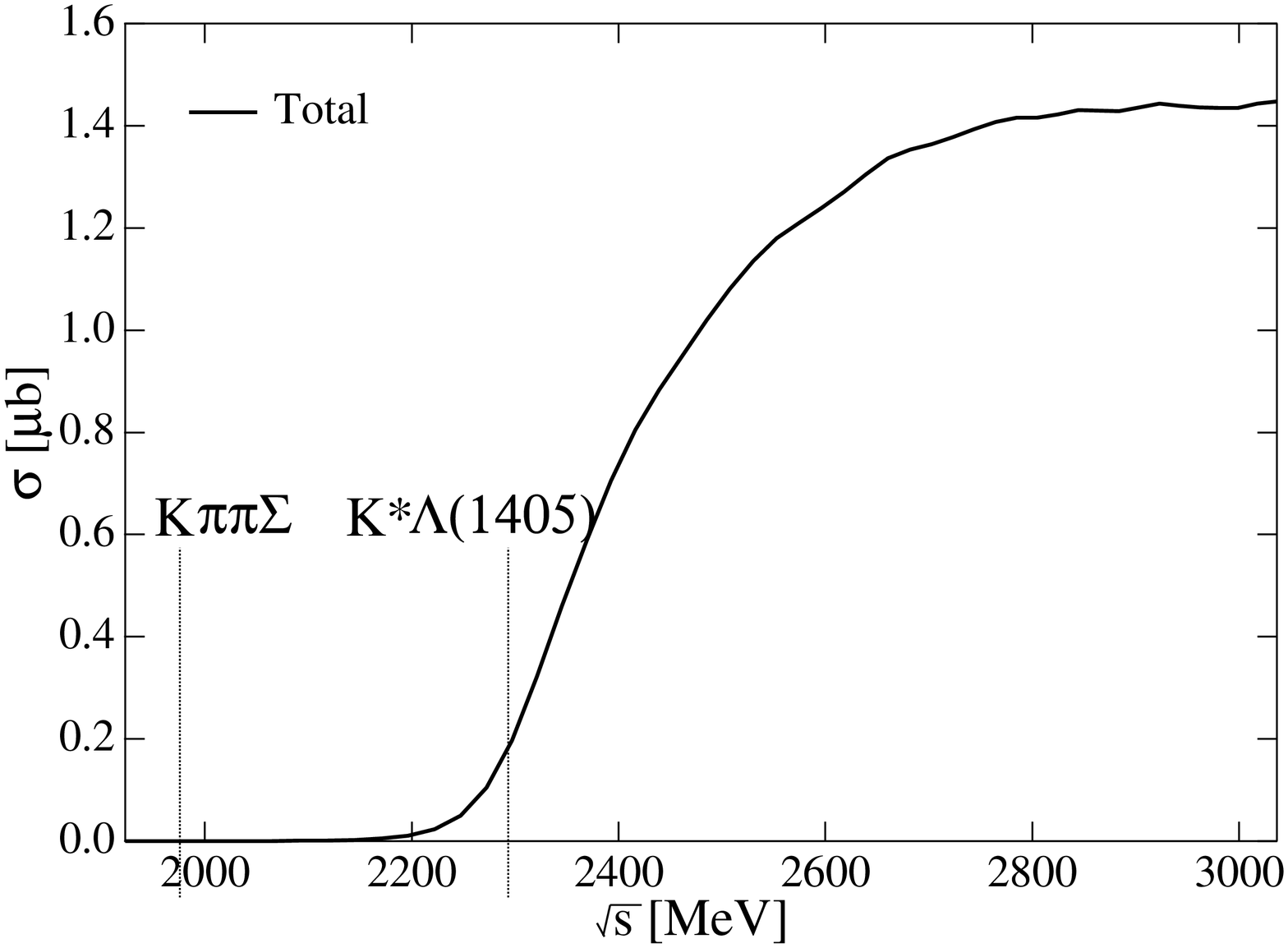}
    \includegraphics[width=8cm,clip]{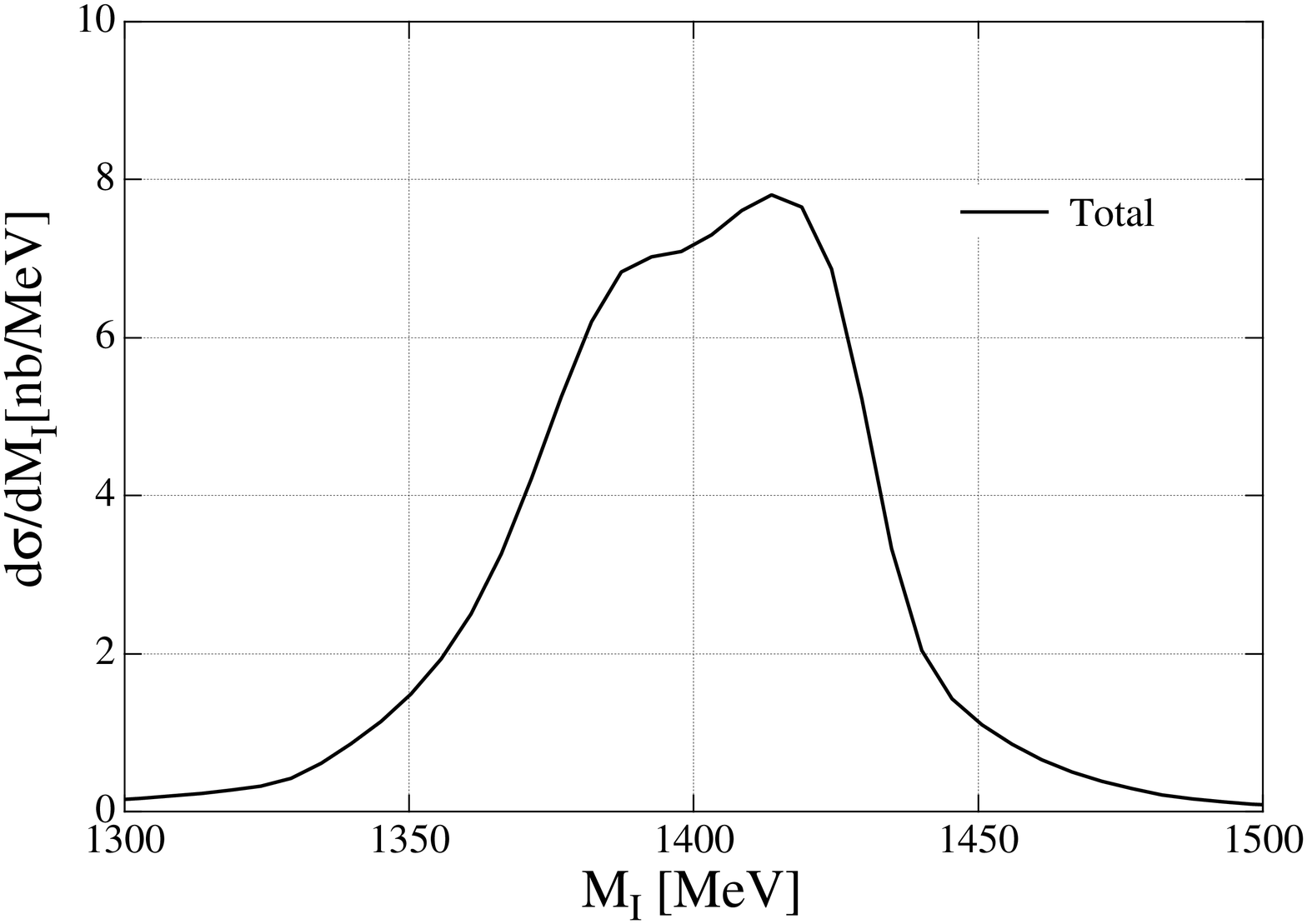}
    \caption{Total cross section and invariant mass distribution
    for the sum of $\pi\Sigma$ and $\pi\Lambda$ channels.}
    \label{fig:total}
\end{figure}

Finally we show the results for the sum of all
$\pi\Sigma$ and $\pi\Lambda$ channels in Fig.~\ref{fig:total}.
This corresponds to the most feasible case in experiment in which 
the three pions decaying from $K^*$ are identified.  
In the total spectrum as a function of $M_I$
(right panel),
we find a two-bump structure reflecting 
both the $\Lambda(1405)$ and the $\Sigma(1385)$.
In the actual case, there would be a further contribution
from the $\bar{K}N$ channel, raising at around 1430 MeV which we do not include
in the calculation.
This contribution starts where the mass distribution in Fig. \ref{fig:total} has
already dropped down and therefore will not blur the shape of the distribution.
This is the case in a related reaction studied in Ref.~\cite{Nacher:1998mi}.
This figure is also illustrating because it reveals a large strength in the
region of 1420 MeV, which makes this shape clearly distinct from the one
observed experimentally in the $\pi^- p\to K^0 \pi\Sigma$
reaction~\cite{Thomas:1973uh}
with a neat peak around 1400 MeV. Hence, this measurement is valuable by
itself.
Yet, to get the individual contributions one should measure the channels
shown in Fig.~\ref{fig:Mdist}.
It is interesting to recall that in the chiral model of
Ref.~\cite{Hyodo:2003jw} it
was shown that the $\pi^- p \to K^0 \pi \Sigma$ reaction favoured the lower mass
$\Lambda$ pole.

In this paper, we have proposed a reaction 
$\gamma p \to \pi^+ K^0 MB$ for the study of 
the second pole possibly existing in the $\Lambda (1405)$ 
region.  
This second resonance has been shown to couple more strongly 
to $\bar KN$ than to $\pi \Sigma$ in several chiral models,  
the present reaction is suitable for the isolation 
of this pole.  
Although the coupling to $\Sigma(1385)$ might contaminate 
the pure $\pi \Sigma$ mass distribution from the second 
$\Lambda(1405)$ pole, the resulting total mass distribution 
still maintains a peak structure pronounced around 1420 
MeV with a relatively narrow width.  
The different shape of this mass distribution would be
well differentiated from other experimental data for the 
$\Lambda (1405)$ excitation induced by other reactions, like the 
$\pi^- p \to K^0 \pi\Sigma$ \cite{Thomas:1973uh}
which favors the lowest energy 
pole at 1390 MeV as shown in Ref.~\cite{Hyodo:2003jw}.
A similar mass distribution to the present one was observed 
in the former study of 
$ K^- p \to \gamma \pi \Sigma$~\cite{Nacher:1999ni}, 
where the photon is emitted from the initial state and hence 
the $\Lambda (1405)$ production is also induced by a $K^-$.
Experimental evidence on the existence of such two
$\Lambda^*$ states would provide more information
on the nature of the current $\Lambda(1405)$ and thus new clues to 
understand non-perturbative dynamics of QCD.

\section*{Acknowledgments}
We would like to thank Prof. T.~Nakano for suggesting this work to us.
We also thank  Dr. D. Jido for useful comments and discussions.
This work is supported by the Japan-Europe (Spain) Research
Cooperation Program of Japan Society for the Promotion of Science
(JSPS) and Spanish Council for Scientific Research (CSIC), which
enabled T. H. and A. H. to visit
IFIC, Valencia and M.J.V. V. visit RCNP, Osaka.
This work is also supported in part  by DGICYT
projects BFM2000-1326,
and the EU network EURIDICE contract
HPRN-CT-2002-00311.

\end{document}